\documentclass[twocolumn,showpacs,pre]{revtex4}
\usepackage{amsmath,amssymb}

\usepackage{epsfig,graphicx}

\usepackage{dcolumn}
\usepackage{bm}

\begin{document}
\title{Effective temperatures of a driven, strongly anisotropic Brownian system}

\author{Min Zhang and Grzegorz Szamel}
\email{szamel@lamar.colostate.edu} \affiliation{Department of
Chemistry, Colorado State University, Fort Collins, Colorado
80523, USA.}

\date{\today}


\begin{abstract}
We use Brownian Dynamics computer simulations of a moderately
dense colloidal system undergoing steady shear flow to investigate
the uniqueness of the so-called effective temperature.
We compare effective temperatures calculated from the fluctuation-dissipation
ratios and from the linear response to a static, long wavelength, external
perturbation along two directions:
the shear gradient direction and the vorticity direction. At high shear rates,
when the system is strongly anisotropic, the fluctuation-dissipation
ratio derived effective temperatures are approximately wave-vector independent,
but the temperatures along the gradient direction are somewhat
higher than those along the vorticity direction.
The temperatures derived from the static linear response
show the same dependence on the direction as those derived from the
fluctuation-dissipation ratio. However, the former and the latter temperatures
are different. Our results suggest that the presently used formulae for
effective temperatures may not be applicable for
strongly anisotropic, driven systems.
\end{abstract}

\maketitle

\section{Introduction}

Temperature is one of the most important parameters describing a
macroscopic system in equilibrium.
In contrast, out of equilibrium, the very notion of temperature,
its definition and its properties
are still a subject of debate. This is true even for stationary nonequilibrium
systems considered in this contribution. In principle, we can always define temperature
in terms of average kinetic energy. It is not clear, however, whether such
a definition leads to a parameter that has a more general significance. In addition,
such a definition cannot be used for systems with overdamped stochastic dynamics
(\textit{e.g.} Brownian dynamics) \cite{Tconf}.

In a seminal paper \cite{CKP} Cugliandolo, Kurchan and Peliti proposed to
define an effective, out-of-equilibrium temperature using a
violation of a fluctuation-dissipation theorem. The starting
point of their analysis was an earlier study \cite{CK} of the
violation of the fluctuation-dissipation theorem in the low temperature
phase of the spherical \textit{p}-spin spin glass system.
Building upon this study
the authors of Ref. \cite{CKP} showed that the ratio of
the correlation and response functions, which in thermal equilibrium
is equal to the temperature, has some properties of temperature even
in out-of equilibrium states. In particular, they showed that an effective
temperature defined through the ratio of the correlation and response functions
controls the direction of the heat flow.

Inspired by Ref. \cite{CKP}, Barrat, Berthier and Kurchan
\cite{BerBarKur,BarBer,BerBar} investigated
the effective temperature defined in terms of the violation of the
fluctuation-dissipation theorem for a classic glass-forming system,
the so-called Kob-Andersen \cite{AndKob} binary Lennard-Jones mixture, undergoing
a steady shear flow. A  stationary drive has some technical advantages
compared to earlier studies of the effective temperature for the same system undergoing
aging \cite{KobBarrat}. The main one is that it can be expected that the system reaches
a stationary non-equilibrium state in which time-translational invariance is restored.
This fact makes easier both a computer simulation study and
a theoretical investigation.
A very interesting feature of a system consisting of particles, as opposed
to mean-field spin systems, is that a great variety of response and correlation
functions and the corresponding fluctuation-dissipation ratios can be defined.
In principle, these different ratios could lead to a diverse set of
effective temperatures. Then, it would be difficult to argue that any
of these temperatures has a deeper meaning. However,
Berthier and Barrat \cite{BerBar} showed that a class of fluctuation-dissipation
ratios results in the same effective temperature. Furthermore, they showed that
the same temperature is registered by a ``thermometer'', \textit{i.e.}
a subsystem coupled to the simulated system.

In a related study Ono \textit{et al.} \cite{Liu1} investigated a model sheared athermal
system (zero-temperature foam). They considered a set of quantities that  are equal to
the temperature
in equilibrium. Some of these quantities
were based on static rather than dynamic linear response relations, and thus they
were different from fluctuation-dissipation ratios
considered in Refs. \cite{CKP,CK,BerBarKur,BarBer,BerBar} from a fundamental
point of view. Ono \textit{et al.} found that
the quantities they considered predicted the same value of effective temperature.
Thus, these quantities could be used to define the effective temperature.

In a follow-up study O'Hern \textit{et al.} \cite{Liu2} addressed
the question of the relation between effective temperatures defined on
the basis of static linear response relations \cite{Liu1} and the fluctuation-dissipation
ratios \cite{BerBarKur,BarBer,BerBar}. They showed explicitly that in \emph{most}
cases these different definitions agreed over two and a half decades of effective
temperature. However, in some cases they obtained different effective temperatures.
The authors of Ref. \cite{Liu2} noted that it is not entirely clear
whether a given pair of conjugate response and correlation function
could be used to define the effective temperature and which definition
of the effective temperature (static or dynamic) should be used.

We briefly mention here a completely different system considered
by Hayashi and Sasa \cite{HS}: a single Brownian particle moving in a tilted
periodic potential. The advantage of this simple system is that many
quantities can be calculated exactly \cite{HS,Reimann}. In particular,
the diffusion coefficient and differential mobility can be calculated and their
ratio can be used to define an effective temperature in close analogy
with fluctuation-dissipation ratios of Refs. \cite{CK,CKP,BerBarKur,BarBer,BerBar}.
Interestingly, Hayashi and Sasa showed that the effective temperature obtained
from the ratio of the diffusion coefficient and differential mobility
plays the role of the usual temperature in the large scale description of
their system. Consequently, they showed that, for this simple system, the
fluctuation-dissipation ratio-derived temperature coincides with the temperature obtained
from a long wavelength limit of a static linear response relation.

Many-particle driven systems for which various definitions of
effective temperatures investigated so far have two common features.
The first one, emphasized for example in Refs.  \cite{CKP,BerBarKur,BarBer,BerBar},
is a separation of time scales: degrees of freedom evolving on short time
scales seem to be thermalized with a different temperature than
those evolving on the longest time scales. The latter degrees of freedom
are thermalized with the effective temperature defined \textit{via}
the fluctuation-dissipation ratio. It should be noted that the
importance of the time scale separation is somewhat less clear for the
static definitions of the effective temperature.
The second common feature is that driven systems considered in Refs.
\cite{BerBarKur,BarBer,BerBar,Liu1,Liu2} were structurally isotropic.
In particular, shear rates used in Ref. \cite{BerBar} were small
enough for the steady state static structure factors in the flow, velocity gradient and
vorticity directions to be identical within simulational uncertainty.
Thus, there was no reason to expect that the effective temperature
defined using wave-vector-dependent correlation functions would depend on the
direction of the wave-vector and thus be anisotropic.

Here we consider a moderately dense model colloidal system undergoing a shear flow
in the limit of high shear rates. As a result of the high shear rate
our system is strongly anisotropic.
In particular, the static structure factor in the velocity gradient direction
is considerably larger than its equilibrium value \cite{Grest} although
no transition to a layered state \cite{Rastogi} is observed. We test three definitions
of effective temperature along two different directions, the velocity gradient
direction and the vorticity direction. The first two definitions are based on
the fluctuation-dissipation ratios. Specifically, the
first definition is based on a violation of the Einstein relation between
the friction coefficient and the self-diffusion coefficient, and the second
definition is based on the violation of the fluctuation-dissipation relation
between the transient response describing the change of the tagged particle density
due to an external potential acting on the tagged particle and the self-intermediate
scattering function. The third definition is a static one based on the violation
of the linear response relation describing the change of the density profile
due to a static, long wavelength, external potential.

All three definitions allow
us to calculate effective temperatures pertaining to different directions relative
to the shear flow. In addition, the second definition allows
us to calculate effective temperatures corresponding to different wave-vectors.
We find that the effective temperatures derived from the fluctuation-dissipation
ratios are wave-vector independent, but at high shear rates the temperatures
along the velocity gradient direction are somewhat larger than
those along the vorticity direction.

The dependence of the effective
temperatures on the direction correlates with the fact that
at high shear rates the system is strongly anisotropic.
In addition, we find that the effective temperatures
obtained from the static linear response relation show the same dependence on the
direction as those derived from the fluctuation-dissipation
ratios. However, the effective temperatures
obtained from the static linear response relation are different from those
derived from the fluctuation-dissipation ratios. Finally, we
show that if the static linear response definition is generalized to
shorter wavelengths (\textit{i.e.} to more rapidly varying in space periodic potentials),
the resulting effective temperatures
have complicated wave-vector and direction dependence.

The paper is organized as follows. In the next section, Sec. \ref{sim},
we describe the details of the simulation. We discuss the structure of
the sheared system in Sec. \ref{struc};
in particular, we show
that at high shear rates the system is strongly anisotropic.
In Secs. \ref{Einstein}, \ref{FDT} and \ref{static} we present effective temperatures
determined using the three different definitions.
We close the paper with a short discussion presented in Sec. \ref{disc}.
In the appendix we examine the wave-vector dependence of the effective
temperatures defined through the static linear response relation.

\section{\label{sim} Simulation details}

We simulated one of the systems investigated in an earlier publication \cite{GS}.
Briefly, the system consists of $N=1372$ spherical colloidal particles.
We used the screened Coulomb potential originally introduced by
Rastogi {\it et al} \cite{Rastogi}:
\begin{equation}\label{potential}
V(r) = A \frac{\exp\left(-\kappa(r-\sigma)\right)}{r},
\end{equation}
with $A=475 k_B T \sigma$
and $\kappa\sigma=24$.
The cutoff distance was $3\sigma$.
We simulated the system at dimensionless densities,
$(N/V)\sigma^3 \equiv n\sigma^3$ equal to 0.408 ($V$ is the system's volume).
We previously established \cite{GS} that this density
corresponds to an effective hard-sphere volume fraction equal to 0.43.
Thus, our system is moderately dense.

We performed Brownian dynamics simulations.  The equation of
of motion for the $i$th particle is,
\begin{equation}\label{Lang}
\dot{\mathbf{r}}_{i} =
\dot{\gamma}y_i\hat{\mathbf{e}}_{x}  - \frac{1}{\xi_0}
\partial_{\mathbf{r}_i}  \sum_{j \ne i}
V\left(\left|\mathbf{r}_i - \mathbf{r}_j \right| \right)
+\mathbf{\eta}_i(t)
\end{equation}
where $\dot{\gamma}$ is the shear rate, $y_i$ is the coordinate of particle $i$
along $y$ direction, $\hat{\mathbf{e}}_{x}$ is the unit vector
along the $x$ direction, $\xi_0$ is the friction coefficient of an isolated particle,
and the random noise $\mathbf{\eta}_i$ satisfies the fluctuation-dissipation
theorem,
\begin{equation}\label{fd}
\left\langle \mathbf{\eta}_i(t) \mathbf{\eta}_j(t') \right\rangle =
2 D_0 \delta(t-t') \delta_{ij} I.
\end{equation}
In equation (\ref{fd}), $I$ is the unit tensor and the diffusion coefficient
$D_0 = k_B T/\xi_0$ where
$k_B$ is Boltzmann's constant.  We will present the results in terms of reduced units
with $\sigma$
and $\sigma^2/D_0$ being the units of length
and time, respectively.

The equations of motion, Eq.~(\ref{Lang}), were solved with Lees-Edwards
boundary conditions for shear rates $\dot{\gamma}=2$, 5, 10 and 20.
To solve the equations of motion we used a second order Brownian dynamics
algorithm \cite{Heun}
with a time step of $10^{-4}$ for shear rates 2 and 5, and a time
step of $5\times 10^{-5}$ for shear rates 10 and 20.

\section{\label{struc} Structure of the sheared system}

A fluid's structure is usually discussed in terms of its pair distribution
function. Under shear, the pair distribution function $g(\mathbf{r})$,
\begin{equation}\label{rdf}
g(\mathbf{r}) = \frac{V}{N(N-1)} \left<\sum_{i \neq j}^N
\delta(\mathbf{r}-\mathbf{r}_j+\mathbf{r}_i)\right>
\end{equation}
is anisotropic (hereafter the brackets $\left< ... \right>$ denote
the non-equilibrium steady state average).  The anisotropic pair
distribution is somewhat difficult to visualize and usually only its
two lowest order projections on spherical harmonics are monitored
\cite{projections}:
\begin{equation}\label{rdfiso}
g^s(r) = \frac{1}{4\pi} \int d\hat{\mathbf{r}} g(\mathbf{r}), \;\;\;
g_+(r) = \frac{15}{4\pi} \int d\hat{\mathbf{r}}\hat{x}\hat{y} g(\mathbf{r}).
\end{equation}
Here we follow notation introduced by Hess and Hanley \cite{HH}:
$g^s(r)$ denotes the isotropic component of the pair distribution function, which
in the absence of the shear flow reduces itself to the familiar equilibrium pair
distribution function $g_{eq}(r)$. Next, $g_+(r)$ denotes the
projection of the pair distribution function onto
$\hat{x}\hat{y}$. In principle, the anisotropic pair distribution $g(\mathbf{r})$
can be reconstructed from the projections
\begin{equation}\label{rdfrec}
g(\mathbf{r}) = g^s(r)  + g_+(r) \hat{x}\hat{y} + \dots,
\end{equation}
but, in practice, for high shear rates
the expansion indicated above is slowly convergent and the
higher order terms, indicated by dots in Eq. \eqref{rdfrec}, are important
\cite{HRH}.

In the limit of low shear rates
$g^s=g_{eq}+o(\dot{\gamma})$,
$g_+$ is proportional to $\dot{\gamma}$ and
all projections other than $g^s$ and $g_+$ are $o(\dot{\gamma})$.
In addition to being the dominant small $\dot{\gamma}$ correction to
the equilibrium pair distribution, the importance of $g_+$ comes from the
fact that for pair-wise additive interactions it determines the $xy$ component
of the interaction contribution to the stress tensor, $\sigma_{xy}$,
\begin{equation}\label{sxy}
\sigma_{xy}= \frac{n^2}{2} \frac{4\pi}{15} \int_0^{\infty} dr r^3 \frac{dV(r)}{dr} g_+(r)
\end{equation}
and, as a consequence, the non-linear viscosity
$\eta(\dot{\gamma})=\sigma_{xy}/\dot{\gamma}$.

\begin{figure}
\includegraphics[scale=0.31]{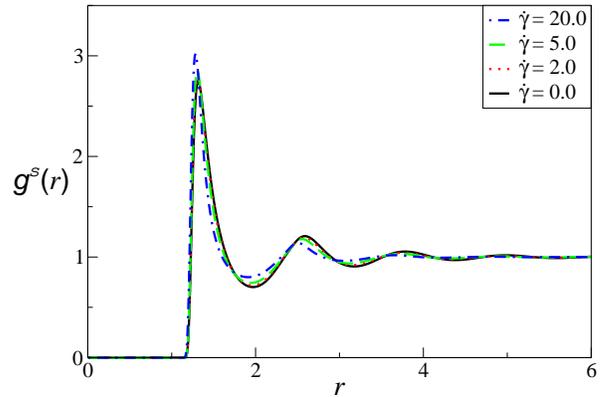}
\caption{\label{f:gr} (Color online) 
The isotropic component of the pair distribution function,
$g^s(r)$, as a function of the particle separation, $r$, for
different shear rates. \\[2ex] }
\end{figure}

\begin{figure}
\includegraphics[scale=0.31]{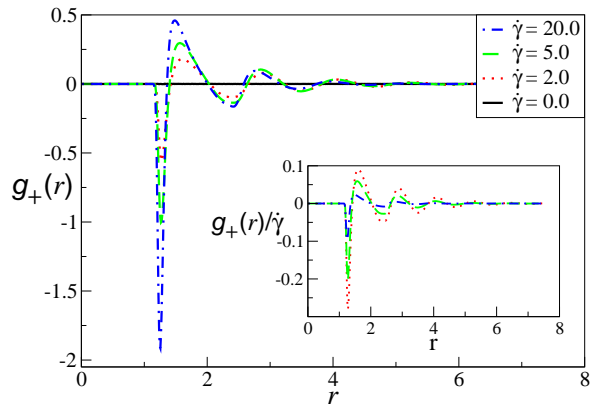}
\caption{\label{f:fxy} (Color online) 
The first anisotropic component of the pair distribution function,
$g_+(r)$, as a function of the particle separation, $r$, for
different shear rates. The inset shows $g_+(r)/\dot{\gamma}$.}
\end{figure}

We show $g^s$ and $g_+$ in Figs. \ref{f:gr} and \ref{f:fxy}. With increasing shear rate
the isotropic component, $g^s$, changes little whereas the amplitude of $g_+$
systematically increases. The inset in Fig. \ref{f:fxy} shows that this increase
of $g_+$ is sub-linear in $\dot{\gamma}$. This fact is a reflection of
the well-known shear thinning phenomenon:
sub-linear increase of $g_+$ leads to a sub-linear
increase of the $xy$ component of the stress tensor and a decrease of the non-linear
viscosity with increasing shear rate.

While the isotropic component of the pair correlation function $g^s$ changes little
with increasing shear, at the largest shear rate investigated, $\dot{\gamma}=20$,
the first anisotropic component $g_+$ is comparable to $g^s$. In fact, although
it is not evident from Fig.  \ref{f:fxy},
the system is strongly anisotropic already at $\dot{\gamma}=5$.
The first evidence of this anisotropy can be obtained by investigating the static
structure factor $S(k)$,
\begin{equation}\label{Sofk}
S(\mathbf{k})=\frac{1}{N}\left<\rho(\mathbf{k})\rho(-\mathbf{k})\right>,
\;\;\; \rho(\mathbf{k}) = \sum_{i=1}^{N}\exp(-i\mathbf{k}\cdot\mathbf{r}_{i}),
\end{equation}
along specific directions in $\mathbf{k}$ space.

\begin{figure}
\vskip 8ex
\includegraphics[scale=0.31]{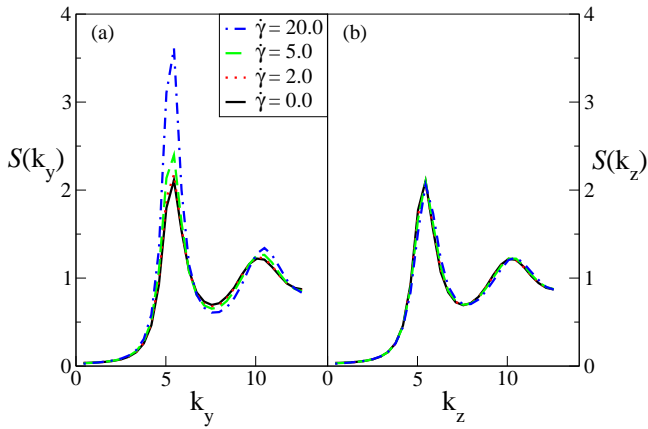}  
\caption{\label{f:sk1d} (Color online)
The structure factor, $S(\mathbf{k})$, for $\mathbf{k}$
along two different directions for different shear rates. 
(a) gradient direction, $\hat{\mathbf{e}}_y$.
(b) vorticity direction, $\hat{\mathbf{e}}_z$.
\\[2ex]}
\end{figure}

In Fig. \ref{f:sk1d} we show the structure factor along two different directions:
the velocity gradient direction $\hat{\mathbf{e}}_y$
and the vorticity direction $\hat{\mathbf{e}}_z$.
We see that the first peak of the structure factor along the gradient direction increases
strongly with shear rate starting at a shear between $\dot{\gamma}=2$ and
$\dot{\gamma}=5$. In contrast, the structure factor along the vorticity direction
shows almost no shear rate dependence. Similar behavior was seen before for a system
under oscillatory shear flow \cite{XueGrest}. The increase of the structure factor
along the velocity gradient direction can be
interpreted in terms of local ordering into a layered-like structure \cite{Rastogi}.

\begin{figure}
\includegraphics[width=8cm, clip=true, trim=4 10 -180 -90]{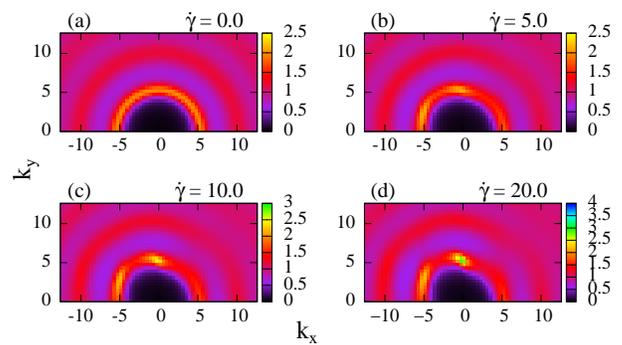}
\caption{\label{f:skxy} (Color online) The structure factor, $S(\mathbf{k})$, in
the velocity-gradient plane, for different shear rates.}
\end{figure}

\begin{figure}
\includegraphics[width=8cm, clip=true, trim=4 10 -180 -90]{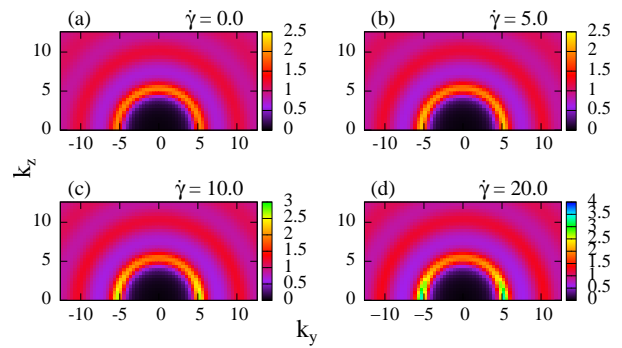}
\caption{\label{f:skyz} (Color online) The structure factor, $S(\mathbf{k})$, in
the gradient-vorticity plane, for different shear rates.}
\end{figure}

To investigate the cross-over between strong shear rate dependence of the
structure factor along the gradient direction and weak shear rate dependence along the
vorticity direction it is useful to visualize
the structure factor in some planes in the wave-vector space.
We investigated the shear rate dependence
of the structure factor in two planes, the flow-velocity gradient plane (x-y) and
the gradient-vorticity (y-z) plane. Fig. \ref{f:skxy} shows that the increase
of the structure factor along the velocity gradient direction is quite localized
in the flow-velocity gradient plane. Moreover, we note that at the highest
shear rate the whole ring-like pattern becomes strongly deformed.
In contrast, Fig. \ref{f:skyz} shows that
in the velocity gradient-vorticity plane the transition from
strong increase of the structure along the gradient direction to almost
shear rate independent structure factor along the vorticity direction is
more gradual. In addition, in the velocity gradient-vorticity plane
the ring-like structure is still quite un-deformed
even at the highest shear rate.

To summarize, in this section we showed that at high shear rates
the steady-state system is strongly anisotropic.

\section{\label{Einstein} Effective temperature:
Einstein relation}

According to the celebrated Einstein relation, the mean-square displacement
of the tagged particle is related to the tagged particle motion under the
influence of a weak external potential acting on the tagged particle only.
Specifically, for a system in equilibrium,
in the long time limit we can define the equilibrium self-diffusion
coefficient $D$ via the tagged particle mean square displacement,
\begin{equation}\label{msdiso}
\lim_{t\to\infty} \frac{1}{t}
\left< \left|\mathbf{r}_i(t)-\mathbf{r}_i(0)\right|^2 \right>_{eq}
= 6 D_{eq},
\end{equation}
where the brackets $\left< ... \right>_{eq}$ denote the equilibrium average.
In addition, we can define the
equilibrium self-mobility coefficient $\mu$ via the systematic
displacement of the tagged particle under the influence of the weak external
force of magnitude $F_0$ acting along an arbitrary direction specified
by the unit vector $\hat{\mathbf{f}}$,
\begin{equation}\label{mobiso}
\lim_{F_0\to 0} \lim_{t\to\infty} \frac{1}{tF_0}
\left< \left(\mathbf{r}_i(t)-\mathbf{r}_i(0)\right)\cdot\hat{\mathbf{f}}
\right>_{eq}^{F_0}
= \mu_{eq},
\end{equation}
where $\left< ... \right>_{eq}^{F_0}$ denotes an average in a system without shear,
subjected to a constant force acting on the tagged particle.

$D_{eq}$ and $\mu_{eq}$ are connected via the so-called Einstein relation
involving the system's temperature $T$,
\begin{equation}\label{Einsteineq}
D_{eq}=T\mu_{eq}.
\end{equation}
For a steady state system under shear both the self-diffusion coefficient
and the self-mobility coefficient depend on the shear rate.
We note here that, for colloidal systems under oscillatory shear,
the shear rate dependence of the former was studied over two decades ago
in experiments \cite{Qiu} and computer simulations \cite{XueGrest}.
These studies were followed by theoretical studies of self-diffusion in
semi-dilute \cite{SBL} and concentrated \cite{Indrani} colloidal systems
under steady shear.
In the former work, the self-mobility was also studied and
the violation of the Einstein relation (\ref{Einsteineq})
was noted. The realization that Eq. (\ref{Einsteineq}) is violated
was not followed, however, by the crucial step of the identification of an
effective temperature. Berthier and Barrat \cite{BerBar,Szamel}
made this last step, turned
Eq. (\ref{Einsteineq}) around, defined and calculated
the effective temperature
\begin{equation}\label{Einsteiniso}
T_{eff}=D/\mu.
\end{equation}
It should be
emphasized that both quantities at the right-hand-side of Eq. (\ref{Einsteiniso})
depend on the shear rate $\dot{\gamma}$ but in a slightly different way. It is this
last fact that gives rise to a non-trivial effective temperature.

Since our system is anisotropic, we investigated two different
effective temperatures based on the violation of Eq. (\ref{Einsteineq})
along two different directions, the gradient direction and the vorticity direction,
\begin{equation}\label{Einsteinyy}
T_{eff}^y=D_{yy}/\mu_{yy},
\end{equation}
\begin{equation}\label{Einsteinzz}
T_{eff}^z=D_{zz}/\mu_{zz}.
\end{equation}
Here $D_{yy}$ and $D_{zz}$ are the diagonal components of the self-diffusion
tensor along the gradient and vorticity directions, respectively,
\begin{equation}\label{msdaa}
\lim_{t\to\infty} \frac{1}{t}
\left< \left(\alpha_i(t)-\alpha_i(0)\right)^2 \right>
= 2 D_{\alpha\alpha},
\end{equation}
where $\alpha=y$ or $\alpha=z$.
Furthermore, in Eqs. (\ref{Einsteinyy}-\ref{Einsteinzz}) $\mu_{yy}$ and
$\mu_{zz}$ are the diagonal elements of the self-mobility tensor
\begin{equation}\label{mobaa}
\lim_{F_0\to 0}\lim_{t\to\infty} \frac{1}{tF_0}
\left< \left(\alpha_i(t)-\alpha_i(0)\right) \right>^{F_0}
= \mu_{\alpha\alpha},
\end{equation}
where, again, $\alpha=y$ or $\alpha=z$, and the weak external force of
magnitude $F_0$ is applied on the tagged particle in the direction $\alpha$.

\begin{figure}
\includegraphics[scale=0.31]{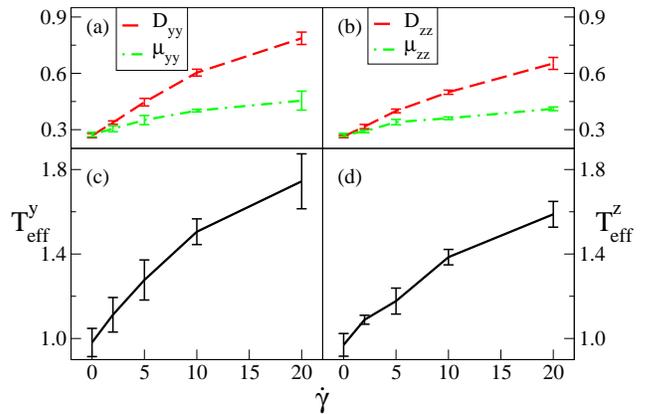}
\caption{\label{f:Einstein} (Color online) (a-b): shear rate dependence of the
diagonal components of the self-diffusion and mobility tensors.
Note that in equilibrium, \textit{i.e}
for $\dot{\gamma}=0$, for our value of temperature, $T=1$, the self-diffusion
and the mobility coefficients are equal, $D_{eq}=\mu_{eq}$.
(c-d): shear rate dependence of the
effective temperature obtained from the self-diffusion and mobility tensors.
(a) and (c): gradient direction, $\hat{\mathbf{e}}_y$. 
(b) and (d): vorticity direction, $\hat{\mathbf{e}}_z$.
}
\end{figure}

The elements of the self-diffusion tensor can be easily calculated. In contrast,
an evaluation of the elements of the self-mobility tensor is a little more challenging.
We followed the procedure described in Ref. \cite{BerBar} and introduced a constant
force acting on 686 randomly selected particles of the system, with the force
in the positive direction along the relevant axis for 343 particles and the force in the
negative direction for the remaining 343 particles. The average systematic
displacement of these particles, weighted by the sign of the force, is a good
approximation for $\left< \left(\alpha_i(t)-\alpha_i(0)\right) \right>$ and
can be used to calculate $\mu_{\alpha\alpha}$. We checked this assertion by
re-doing this calculation with a constant force acting on 340 randomly selected
particles of the system. As expected, the statistics was worse, but the results
were compatible with the calculation using a constant force acting on 686 particles.

The results are presented in Fig. \ref{f:Einstein}. The upper panels
of this figure show that both the diagonal elements of the self-diffusion tensor
and the diagonal elements of the self-mobility tensor increase with increasing shear
rate. Both quantities are somewhat larger along the gradient direction than
along the vorticity direction. We note that for a system under an oscillatory shear
an early simulation of Xue and Grest \cite{XueGrest} found the opposite behavior
for the elements of the self-diffusion tensor. In contrast, for a system under
steady shear a theoretical study of Indrani and Ramaswamy \cite{Indrani}
found that the diagonal elements of the self-diffusion tensor along the gradient
direction was larger than those along the vorticity direction. This
qualitatively agrees with the results for the elements of the self-diffusion tensor
showed in Fig. \ref{f:Einstein}. It should
be emphasized, however, that in Ref. \cite{Indrani} the violation of the
Einstein relation was neglected and thus Indrani and Ramaswamy's approach
cannot describe the difference between self-diffusion and self-mobility
tensors.

The lower panels of Fig.
\ref{f:Einstein} show that, as found in earlier investigations, the
effective temperature increases with increasing shear rate. The temperatures
along the gradient and vorticity directions are quite close, but the
one along the gradient direction is systematically somewhat larger than
the one along the vorticity direction. We should note, however, that the
uncertainty of these effective temperatures prevents us from making a definitive
statement that the two temperatures are different.

\section{\label{FDT} Effective temperature: FDR
involving the self-intermediate scattering function}

According to the standard fluctuation-dissipation relation,
in equilibrium the susceptibility describing a
time-delayed change of the tagged particle density
is proportional to the self-intermediate scattering function,
\begin{equation}\label{Fskteq}
\chi_{s,eq}(k;t) = \frac{1}{T} \left[ 1 - F_{s,eq}(k;t)\right].
\end{equation}
Here $\chi_{s,eq}(k;t)$ is the susceptibility describing the change of the
Fourier component $k$ of the tagged particle density due to a weak external
force acting on the tagged particle,
\begin{equation}
\chi_{s,eq}(k;t) = \frac{\partial }{\partial h}
\left< \exp\left[i\mathbf{k}\cdot\mathbf{r}_i(t)\right] \right>^{h}_{eq}\,_{|_{h=0}}
\end{equation}
where $\left< ... \right>^{h}_{eq}$ denotes an average calculated in a system that
was in equilibrium at $t=0$ and subsequently was subjected to
a constant in time force $\mathbf{F}_i^{ext}$ acting on the
tagged particle only,
$$
\mathbf{F}_i^{ext} = -\partial_{\mathbf{r}_i} \left[
2 h \cos\left(\mathbf{k}\cdot\mathbf{r}_i
\right) \right].
$$
Furthermore, $F_{s,eq}(k;t)$ in Eq. \eqref{Fskteq} is the equilibrium
self-intermediate scattering function,
\begin{equation}
F_{s,eq}(k;t) = \left<
\exp\left[i\mathbf{k}\cdot\left(\mathbf{r}_i(t)-\mathbf{r}_i(0)\right)\right]\right>.
\end{equation}

For a system undergoing a steady shear flow, the relation \eqref{Fskteq}
is violated. Following \cite{BerBar} we define an effective temperature through
the violation of this relation,
\begin{equation}\label{Fskt}
T^{\alpha}_{eff} = \frac{1 - F_s(k_{\alpha};t)}{\chi_s(k_{\alpha};t)}
\end{equation}
All quantities at the right-hand-side of Eq. \eqref{Fskt}
are calculated for a system undergoing
a steady shear. Specifically, $F_s(k_{\alpha};t)$ is the steady state
self-intermediate scattering function. Moreover,
$\chi_s(k_{\alpha};t)$ describes the
time-delayed change of the tagged particle density,
\begin{equation}\label{chi}
\chi_{s}(k;t) = \frac{\partial }{\partial h}
\left< \exp\left[i\mathbf{k}\cdot\mathbf{r}_i(t)\right] \right>^{h}\,_{|_{h=0}},
\end{equation}
where $\left< ... \right>^{h}$ denotes an average calculated in a system that
was undergoing a steady shear at $t=0$, and subsequently was subjected
to a constant in time force acting on the tagged particle. It should be noted
that the effective temperatures in definition \eqref{Fskt}, in
addition to the dependence on the direction of the
force, $\alpha=y,z$, can also depend on the magnitude of the wave-vector
(to simplify notation this possible dependence is not indicated explicitly).
As discussed in the Introduction, the authors of Ref. \cite{BerBar} found
that, for a supercooled system under weak shear, the effective temperatures
defined through Eq. \eqref{Fskt} does not depend on the wave-vector.

\begin{figure}
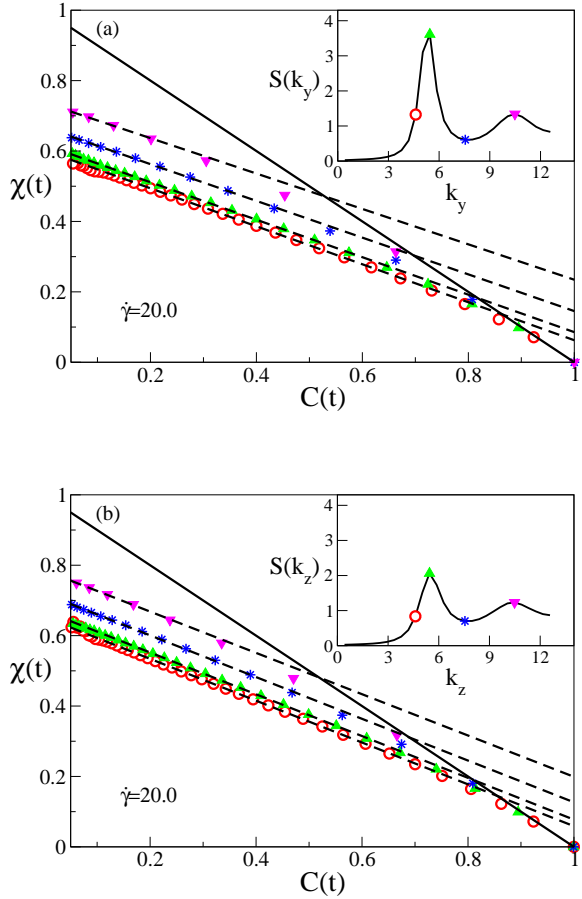

\includegraphics[scale=0.31]{fig7a.eps} \\[7ex]
\includegraphics[scale=0.31]{fig7b.eps}
\caption{\label{f:fdt20} (Color online) 
Parametric plots $\chi_s(k_{\alpha};t)$ vs.
$F_s(k_{\alpha};t)$ for wave-vectors along two directions. (a) gradient
direction, $\hat{\mathbf{e}}_y$. (b) vorticity direction, $\hat{\mathbf{e}}_z$. 
The magnitudes of the wave-vectors are
$k_{\alpha} = 4.61$ 5.45, 7.55 and 10.48, listed from bottom to top.
The solid lines have slopes equal to $-1/T=-1$.
The dashed lines have slopes equal to $-1/T_{eff}^{\alpha}$.
The insets indicate the magnitudes
of the wave-vectors in relation to the steady state structure factors
for wave-vectors along the gradient and vorticity directions.
Recall that at this high shear
rate the steady state structure factors along these directions
are quite different, see Fig. \ref{f:sk1d}.}
\end{figure}

The calculation of the self-intermediate scattering function is straightforward.
In contrast, it is more difficult to calculate the susceptibility. We followed the
procedure described in Ref. \cite{BerBar}. Specifically, we considered a system
with a constant force imposed on 686 randomly selected particles of the system,
with the force in the positive direction along the relevant axis for 343 particles
and the force in the negative direction for the remaining 343 particles.
Furthermore, in order to evaluate the susceptibility using the un-perturbed
trajectories, we followed a recently introduced approach \cite{Berthier} and
we performed the derivative in Eq. \eqref{chi} \emph{before} taking
the average. This necessitates solving additional equations of motion for
quantities $\partial \mathbf{r}_i/\partial h$ \cite{Berthier},
but it allowed us to calculate the susceptibility from one long, unperturbed trajectory.
It should be noted that, in contrast to the calculation described in Ref.
\cite{Berthier}, we did not have to perform several independent runs. This
simplification follows from the fact that while in Ref. \cite{Berthier}
a response of an aging system was studied, we are interested in a response
of a steady state system.

\begin{figure}
\includegraphics[scale=0.31]{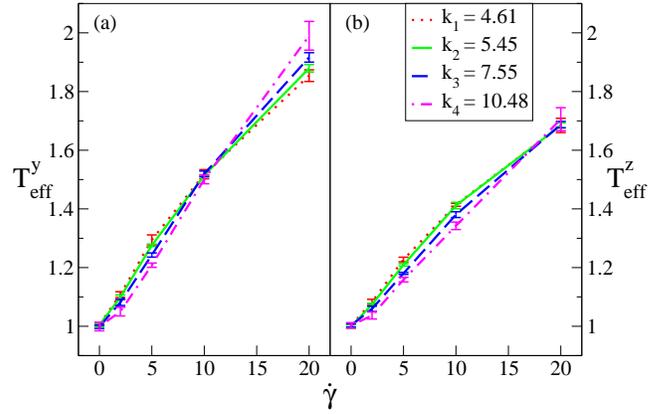}
\caption{\label{f:fdtteff} (Color online) Shear rate dependence of effective
temperatures obtained from the violation of the fluctuation-dissipation relation
involving the self-intermediate scattering function.
(a): gradient direction, $\hat{\mathbf{e}}_y$. 
(b): vorticity direction, $\hat{\mathbf{e}}_z$.
}
\end{figure}

In Fig. \ref{f:fdt20} we show parametric plots of the susceptibility versus
the correlation function for several wave-vectors along the $y$ and $z$ axes
at the highest shear rate, $\dot{\gamma}=20$. In plots of this type the
effective temperature is the (minus) of the inverse slope of the data.
Moreover, short times correspond to the lower right corner and long
times to the upper left corner of the plot.
One can see from Fig. \ref{f:fdt20} that at short times
the fluctuation-dissipation \eqref{Fskteq} is obeyed. In contrast, at long times
an effective temperature higher than the heat bath temperature is observed.
Notice that at long times, for a given direction,
the slopes of the lines fitted to
the data obtained at different magnitudes of the wave-vectors seem very close.
This agrees with the findings of Ref. \cite{BerBar}.
In contrast, the slopes of the lines fitted to
the data obtained at two different directions are different.

The results of the analysis of parametric plots at all shear rates
investigated are presented in Fig. \ref{f:fdtteff}. In agreement with the
behavior found from the analysis of the violation of the Einstein relation,
we find that that in both directions and at all wave-vectors the
effective temperatures increase with increasing shear rate. The temperatures
obtained from different magnitudes of the wave-vectors are reasonably
consistent. This agrees with the finding of Berthier and Barrat \cite{BerBar}.
As found in the previous section, temperatures obtained from the external force
along the gradient and vorticity directions are close, but the
one along the gradient direction is systematically larger than
the one along the vorticity direction.

\section{\label{static} Effective temperature:
static linear response definition}

For a system in equilibrium,
the static structure factor plays the role of the response function:
the change of the average density due to a weak, static,
periodic in space, external potential,
\begin{equation}\label{extpot}
V(r) = V_0 \sin(\mathbf{q}\cdot\mathbf{r}),
\end{equation}
which has been turned on at $t=-\infty$,
is given by the following well known relation:
\begin{equation}\label{denchange}
\delta n(\mathbf{r}) = - \frac{n S_{\mathrm{eq}}(q)}{T}
V_0 \sin(\mathbf{q}\cdot\mathbf{r})
\end{equation}
where $T$ is the system's temperature and
$S_{\mathrm{eq}}(q)$ is the equilibrium structure factor.

The main finding of Ref. \cite{Liu2} is that the effective temperature obtained
from the long wavelength (\textit{i.e.} small wave-vector) limit of \emph{some}
linear response relations agrees with that obtained from the fluctuation-dissipation
ratios. Inspired by this result here we use the relation \eqref{denchange}
to define an effective temperature. Specifically, we assume a sheared, stationary
suspension,
impose a slowly varying external potential \eqref{extpot} (we use the smallest
wave-vector allowed by periodic boundary conditions), wait a time long enough
for the system to reach a new stationary state, and measure the density change
due to the external potential. We use two different strengths of the effective
potential (\textit{i.e.} two different values of $V_0$) to extrapolate to the linear
response limit implicit in Eq. \eqref{denchange}.
In the limit of weak external potential the density change is proportional to
$V_0$,
\begin{equation}\label{denchangelr}
\delta n(\mathbf{r}) = - n a(\mathbf{q})  V_0
\sin(\mathbf{q}\cdot\mathbf{r})
\end{equation}
where we made explicit the fact that under shear the density change depends on the
orientation of the vector $\mathbf{q}$ with respect to the shear flow.
Next, we use the following generalization
of the relation \eqref{denchange} to extract the static effective temperature:
\begin{equation}\label{effTstatic}
T_{\mathrm{eff}}^\alpha =
\frac{S(q_\alpha)}{a(q_\alpha) }
\end{equation}

It should be noted that the definition \eqref{effTstatic}
involves the shear rate dependent, anisotropic structure factor $S(\mathbf{q})$.
This agrees with the static definition adopted in Ref. \cite{Liu2}.
Thus, the static effective temperature \eqref{effTstatic}
is a ratio of two shear rate-dependent quantities.

\begin{figure}
\includegraphics[scale=0.31]{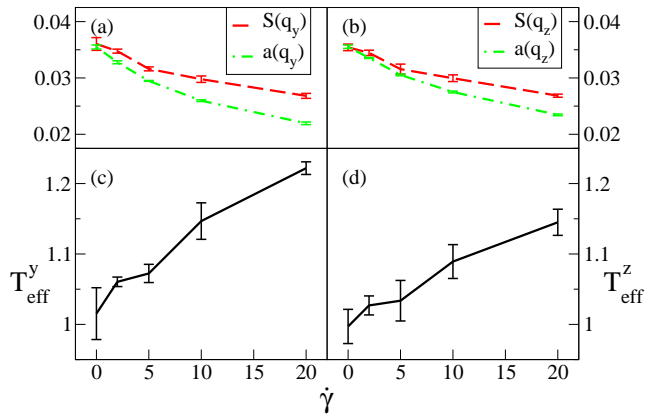}
\caption{\label{f:static0} (Color online) (a-b): shear rate dependence of the
steady state structure factor $S(q_\alpha)$ and the response coefficient $a(q_\alpha)$,
$\alpha=y,z$,  where $q_\alpha$ is the smallest non-zero
wave-vector allowed by the periodic boundary conditions, $q_y=q_z=0.42$.
(c-d): shear rate dependence of the
effective temperature obtained from the static linear response.
(a) and (c): gradient direction, $\hat{\mathbf{e}}_y$. 
(b) and (d): vorticity direction, $\hat{\mathbf{e}}_z$.
}
\end{figure}

In contrast to the procedure adopted in Ref. \cite{Liu2} we did not
monitor the time dependent (transient) density change after the external potential
was turned on, and thus do not obtain the static effective temperature from
an intercept of a parametric plot of the response (\textit{i.e.} the density change)
as a function of the time dependent density correlation function.
However, our definition of the effective temperature
is equivalent to that used in Ref. \cite{Liu2}.

In Fig. \ref{f:static0} we present the shear rate dependence of the
structure factor and the linear response
coefficient $a$. Since we are interested in the response
to a long wavelength perturbation, both quantities are shown for the smallest
wave-vector compatible with periodic boundary conditions, $q_y=q_z=0.42$.
While both quantities decrease with increasing shear rate, the coefficient $a$ has
a somewhat stronger dependence on $\dot{\gamma}$. As a result, as shown in
lower panels in Fig. \ref{f:static0}, the effective temperatures calculated
for the perturbation along the gradient and vorticity directions increase
with shear rate. The effective temperature along the gradient direction is
larger than that along the vorticity direction. This agrees with the behavior
obtained from the Einstein relation and the fluctuation-dissipation ratios.
However, the values of the effective temperature obtained from static linear
response are different than those obtained from the fluctuation-dissipation ratios.

\section{\label{disc} Discussion}

\begin{figure}
\vskip 6ex
\includegraphics[scale=0.31]{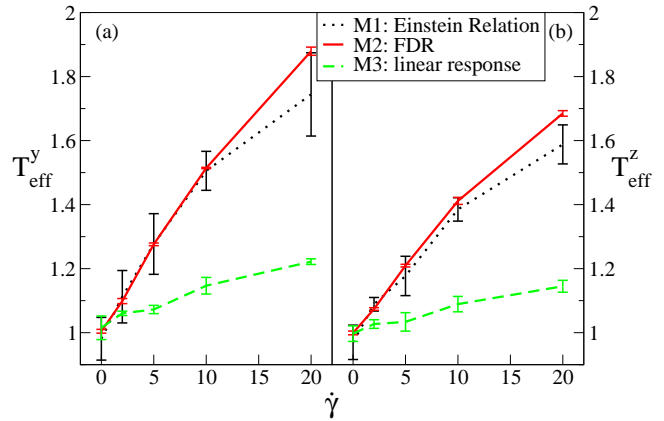}
\caption{\label{f:teffcomp} (Color online) Comparison of the shear rate dependence of
effective temperatures obtained using three different definitions. Dotted line:
Einstein relation; solid line: violation of the fluctuation-dissipation relation
for the self-intermediate scattering function at the wave-vector corresponding to
the peak position of the
structure factor; dashed line: static linear response to a long wavelength
perturbation. (a) gradient direction, $\hat{\mathbf{e}}_y$. 
(b) vorticity direction, $\hat{\mathbf{e}}_z$. 
}
\end{figure}

To investigate
the uniqueness of the notion of the so-called effective temperature of a driven
colloidal system we studied three definitions of such a temperature. The
main results are summarized in Fig. \ref{f:teffcomp}.

We evaluated the effective temperature obtained from the
violation of the Einstein relation \cite{BerBar} and the effective temperature obtained
from the violation of the fluctuation-dissipation relation involving the
self-intermediate scattering function \cite{BerBar}.
The latter definition allows to calculate
the effective temperature at different wave-vectors. Formally, the former
definition can be considered as the limiting case of the latter one for
vanishing magnitude of the wave-vector. Berthier \textit{et al.} \cite{BerBar} found that
these two definitions resulted in compatible values of the effective temperatures.
Our results suggest that at higher shear rates, when the system is anisotropic,
the two definitions give the same result along the gradient direction and
along the vorticity direction. However, the effective temperatures along the
gradient direction are somewhat larger than those along the vorticity direction.

In addition, we investigated a definition of the effective temperature based
on a static linear response relation. The motivation for this was two-fold.
First, similar relations were investigated by O'Hern \textit{et al.} \cite{Liu2}.
Second, Hayashi and Sasa \cite{HS} found that for their simple, single
particle system, the effective temperature
defined through a fluctuation-dissipation ratio determines the response of the
system to a long wavelength potential. We found that the effective temperatures
obtained from the static linear response relation were different from those
obtained from violation of the Einstein relation and of
the fluctuation-dissipation relation involving the
self-intermediate scattering function. However, in the long wavelength limit
static linear response derived effective temperatures had the same dependence on the
direction: temperatures along the gradient direction were somewhat larger
than those along the vorticity direction. In the appendix we extended the
static linear response definition of the effective temperature to
shorter wavelength perturbations. As shown in Fig. \ref{f:teffstatic},
at shorter wavelengths (larger wave-vectors) the static linear response relation
leads to a variety of different effective temperatures, and in some cases with
dependence on the direction opposite to that shown in Fig. \ref{f:teffcomp}.

It would be interesting to repeat our numerical investigation for a more
strongly interacting system with a clear separation of time scales. To this
end, however, one would have to use a mixture (possibly, Kob-Andersen binary
Lennard-Jones mixture) because we found that our single component system
forms layers at high density and large shear rates (as was first reported in
Ref. \cite{Rastogi}).

It would also be interesting to develop some theoretical understanding
of the definitions of the effective temperature investigated here. 
First steps in this direction have already been taken \cite{Szamel,KF}. 
It is still not clear, however, when one has to use a fluctuation-dissipation
based definition of the effective temperature and when one can use a static
linear response based definition for a many-particle, strongly interacting, sheared
system.

More generally, it would be worthwhile to establish what, if any, properties
of a many-particle, strongly interacting, sheared system are influenced by which
definition of the effective temperature. Again, first steps in this direction have 
already been 
taken \cite{HaxtonLiu}. We would like to mention here a very interesting 
result concerning driven systems with Newtonian dynamics: alignment of particles
in a shear flow depends on whether kinetic or configurational \cite{Tconf} 
temperature is thermostated \cite{confthermostat}.  

The main conclusion from our study is that for a strongly driven, anisotropic
system presently used, definitions of the effective temperature may not give
a unique, direction-independent result. It is, therefore, imperative that 
the realms of validity and importance are established for these definitions. 
Only then we will be able to fully assess the relevance of the existing definitions.

\section*{Acknowledgments}
We thank E. Flenner for comments on the manuscript and
gratefully acknowledge the support of NSF Grant No.\ CHE 0909676.

\appendix
\section*{Appendix}

Here we generalize the static linear response based definition of effective temperature.
The main motivation for this generalization
is that, as shown in Fig. \ref{f:teffcomp}, the
static linear response to a long wavelength perturbation gives effective temperatures
that are different from those obtained using the fluctuation-dissipation ratios.
This difference prompted us to check whether it is still present at
larger wave-vectors.

\begin{figure}
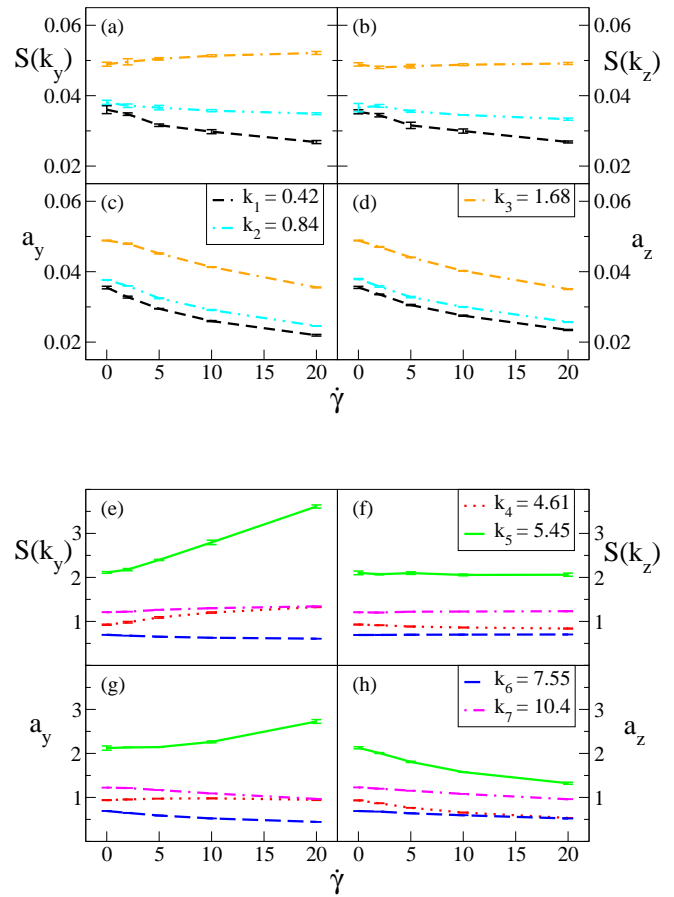

\vskip 8ex
\includegraphics[scale=0.31]{fig11abcd.eps} \vskip 7ex
\includegraphics[scale=0.31]{fig11efgh.eps}
\caption{\label{f:responsestatic} (Color online) 
(a-b) and (e-f): shear rate dependence of the structure factor. 
(c-d) and (g-h): shear rate dependence of the
linear response coefficient. 
(a), (c), (e) and (g): gradient direction, $\hat{\mathbf{e}}_y$.
(b), (d), (f) and (h): vorticity direction, $\hat{\mathbf{e}}_z$.
(a-d) show smaller wave-vectors and (e-h) show larger wave-vectors.
\\[2ex]
}
\end{figure}

\begin{figure}
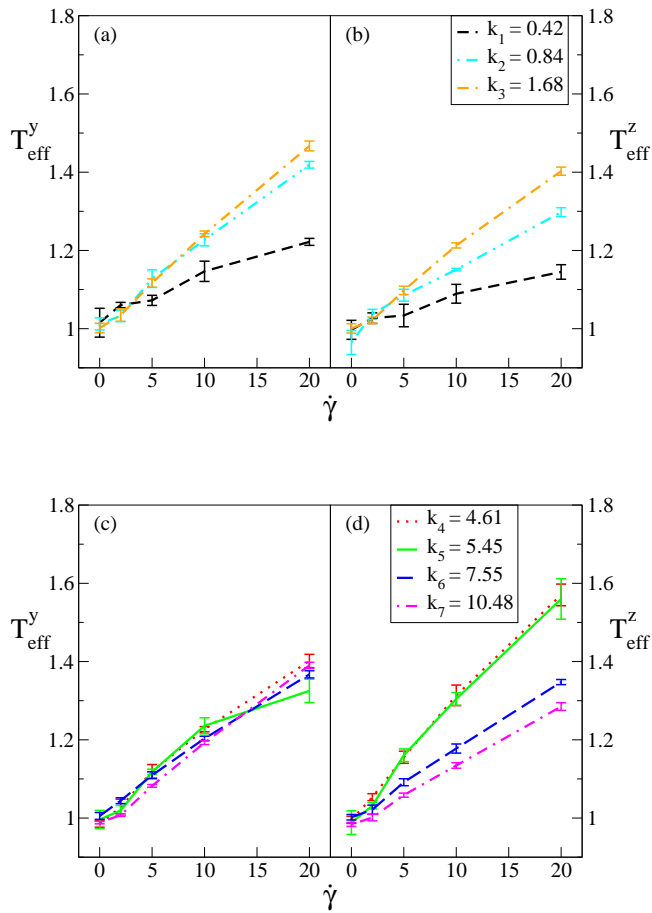

\vskip 7ex
\includegraphics[scale=0.31]{fig12ab.eps} \vskip 7ex
\includegraphics[scale=0.31]{fig12cd.eps}
\caption{\label{f:teffstatic} (Color online) 
Shear rate dependence of the static effective temperature
at a number of wave-vectors. 
(a) and (c) : gradient direction, $\hat{\mathbf{e}}_y$. 
(b) and (d): vorticity direction, $\hat{\mathbf{e}}_z$.
(a-b) show smaller wave-vectors and (c-d) show larger wave-vectors.
}
\end{figure}

Effective temperatures presented in this section were obtained using relation
\eqref{effTstatic} generalized to larger wave-vectors:
\begin{equation}\label{effTstatic1}
T_{\mathrm{eff}}^\alpha =
\frac{S(k_\alpha)}{a(k_\alpha) }
\end{equation}
In Eq. \eqref{effTstatic1} the effective temperature may depend on the wave-vector.

In Figs. \ref{f:responsestatic} and \ref{f:teffstatic} we show the shear
rate dependence of the structure factor, linear response coefficient, and
effective temperature for different wave-vectors along two directions, the
gradient direction and the vorticity direction. We find that effective temperature
as defined through Eq. \eqref{effTstatic1} depends significantly on the wave-vector.
In particular, whereas at small wave-vectors temperatures along the gradient
direction are somewhat larger than those along the vorticity direction, at
some of the larger wave-vectors this ordering is reversed. Our finding
emphasizes the need for some theoretical understanding of the static linear
response definition of effective temperature.


\begin{thebibliography}{99}
\bibitem{Grest} The increase of the steady state structure factor along the
velocity gradient direction was also seen in model suspensions undergoing
oscillatory shear, \cite{XueGrest}.
\bibitem{projections}Higher-order projections were examined by Hess and Hanley \cite{HH}.
\end{thebibliography}
\end{document}